\documentclass[aps,prl,twocolumn,groupedaddress]{revtex4}
\usepackage[pdftex]{graphicx}
\begin{document}
\title{Magnetic field induced charge redistribution in disordered graphene double quantum dots}

\author{K. L. Chiu$^{1,2}$, M. R. Connolly$^{1,3}$, A. Cresti$^{4,5}$, J. P. Griffiths$^1$, G. A. C. Jones$^1$, C. G. Smith$^1$}

\affiliation{$^1$Cavendish Laboratory, Department of Physics, University of Cambridge, Cambridge, CB3 0HE, UK}
\affiliation{$^2$Department of Physics, Massachusetts Institute of Technology, Cambridge, MA 02139, USA}
\affiliation{$^3$National Physical Laboratory, Hampton Road, Teddington TW11 0LW, UK}
\affiliation{$^4$Univ. Grenoble Alpes, IMEP-LAHC, F-38000 Grenoble, France}
\affiliation{$^5$CNRS, IMEP-LAHC, F-38000 Grenoble, France}

\date{\today}

\begin{abstract}
We have studied the transport properties of a large graphene double quantum dot under the influence of background disorder potential and magnetic field. At low temperatures, the evolution of the charge-stability diagram as a function of \textit{B}-field is investigated up to 10 Tesla. Our results indicate that the charging energy of quantum dot is reduced, and hence the size of the dot increases, at high magnetic field. We provide an explanation of our results using a tight-binding model, which describes the charge redistribution in a disordered graphene quantum dot \textit{via} the formation of Landau levels and edge states. Our model suggests that the tunnel barriers separating different electron/hole puddles in a dot become transparent at high \textit{B}-fields, resulting in the charge delocalization and reduced charging energy observed experimentally.

\end{abstract}

\maketitle
\section{I. INTRODUCTION}

Confining charge carriers in graphene continues to generate interest owing to its customizable electronic properties and compatibility with existing semiconductor device processing \cite{Geim2007}. Carbon atoms $^{12}$C have low atomic weight and no nuclear spin (except for the $^{13}$C isotope), so electronic interactions such as spin-orbit and hyperfine coupling are expected to be weak in graphene, leading to long electron-spin relaxation times \cite{Huertas-Hernando2009}. Over the past decade lithographically-defined graphene quantum dots (GQDs) have proved to be an useful platform in which single electrons can be confined and manipulated. A number of experimental advances have been reported, such as charge detection \cite{Gutinger2008}, charge relaxation \cite{Volk2013}, and electron-hole crossover \cite{Guttinger2009} in graphene single quantum dots (GSQDs); and excited states \cite{Molitor2010}, tunable interdot coupling \cite{Liu2010} and charge pumping \cite{Connolly.2013a} in graphene double quantum dots (GDQDs). More recently, graphene quantum dots on hexagonal boron nitride (hBN) have enabled the influence of potential and edge disorder to be studied separately \cite{Engels2013,Epping2013}. Magnetic fields are a powerful tool for unveiling the nature of confined Dirac Fermion in GQDs. For example, the Fock-Darwin spectrum in the few-electron regime \cite{Guttinger2009} and many-electron regime \cite{Chiu2012} as well as the Zeeman splitting of spin states \cite{Guttinger2010} in graphene single quantum dot have been studied. On the other hand, while it is well-known that electron transport through graphene nanostructures is strongly affected by electron/hole puddles induced by potential fluctuations \cite{Liu2010,Volk2011,Amet2012}, detailed experimental and theoretical studies are lacking to address this issue in GQD transport. In this letter, we study the effect of disorder by investigating the transport properties of a large GDQD device at magnetic fields in which Landau levels (LLs) are expected to form. At high enough \textit{B}-field, our results suggest that electron/hole puddles in the dot tend to merge together, giving rise to a charge redistribution which can be observed experimentally. Our results are supported by tight-binding quantum simulations, which can be used to describe the charge redistribution in a disordered graphene quantum dot at high magnetic fields and gives deeper insight into our experimental data.

\maketitle
\section{II. Coulomb blockade measurement on a graphene double dot at B=0}

\begin{figure}[!t]	
\includegraphics[scale=1]{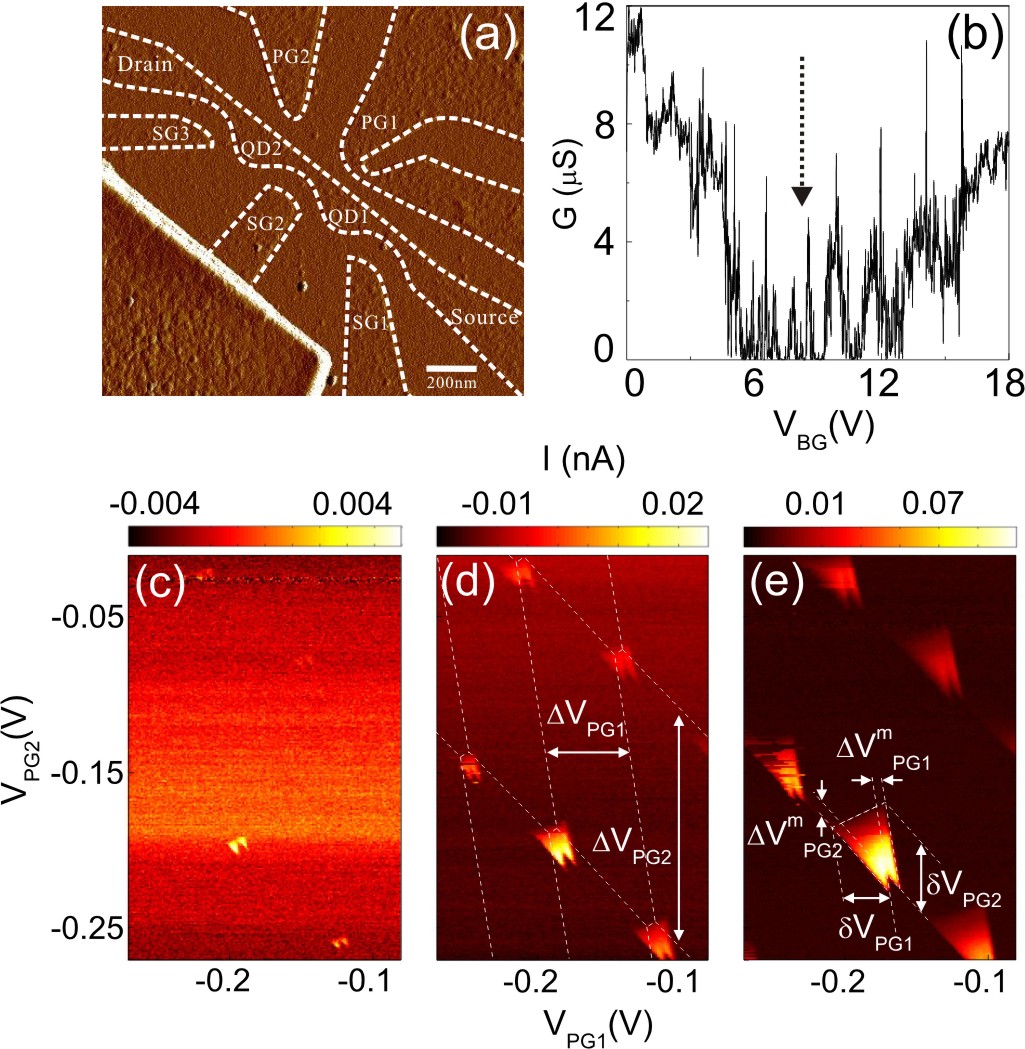}
\caption{(a) Atomic force micrograph of the double quantum dot device measured in this work. (b) Measurement of the differential conductance through the DQD for varied back gate voltages. Data collected at $V_{AC}$=200 $\mu$V and $T$=1.4 K. Current through DQD as a function of $V_{PG1}$ and $V_{PG2}$ measured in a dilution fridge at $T$=100 mK with applied DC bias (c) $V_{SD}$=400 $\mu$V, (d) $V_{SD}$=1 mV and (e) $V_{SD}$=2 mV. The position of the triple points can be determined at low bias (see (c)), while at high bias the triple points evolve into triangles (see (d),(e)).}    
\label{Fig3}
\end{figure}

Double quantum dots are a model system for investigating the dynamics of electrons in a wide range of semiconductors \cite{Hanson2007,Wiel2002,Pfund2006,Schroer2011,Chorley2011,Pecker2013,Zwanenburg2013}. Charge stability diagrams - obtained by measuring the conductance as a function of the carrier density on each quantum dot - reveal a wealth of information about charging energy, interdot coupling and cross gate coupling strength, making them an ideal way to probe charge rearrangments in quantum dots at high magnetic fields. An atomic force microscopy image of the double quantum dot measured in this work is shown in Fig. \ref{Fig3}(a). Our device consists of two lithographically etched (O$_{2}$ plasma) graphene islands each with a size of 200$\times$250 nm$^{2}$ labeled QD$_{1}$ and QD$_{2}$ in Fig. \ref{Fig3}(a). They are mutually connected to each other by a 90 nm wide constriction and separately connected to the source/drain leads \textit{via} two 80 nm wide constrictions, which act as tunnel barriers. Two plunger gates PG1(2) are used to tune the energy levels in QD$_{1(2)}$ while three side gates (SG1, SG2 and SG3) are used to tune the tunnel barriers. The doped-silicon back gate (BG) is used to adjust the overall Fermi level.

The measurements were performed in a dilution refrigerator with an electron temperature around 100 mK. In Fig. \ref{Fig3}(b), we show the measured differential conductance through DQD as a function of BG voltage ($V_{AC}$=20 $\mu$V) highlighting a region of suppressed current (the so-called transport gap \cite{Stampfer2009}) separating the hole from the electron transport regime. At a back gate voltage within the transport gap ($V_{BG}$=8.61 V, see arrow in Fig. \ref{Fig3}(b)) we measure the DC current through DQD as a function of $V_{PG1}$ and $V_{PG2}$ for a series of applied DC biases as shown in Fig. \ref{Fig3}(c) for $V_{b}$=400 $\mu$V, (d) for $V_{b}$=1 mV and (e) for $V_{b}$=2 mV respectively. 
 
As expected, the current in the stability diagram evolves from triple points into bias-dependent triangles when the bias is increased. The horizontal and vertical measure of the honeycomb cell $\Delta$$V_{PG1}$ and $\Delta$$V_{PG2}$ (Fig. \ref{Fig3}(d)) give the capacitances between the gate PG1 and QD$_{1}$ $C_{g1}$$\approx$$e/\Delta$$V_{PG1}$=2.77 aF, and between the gate PG2 and QD$_{2}$ $C_{g2}$$\approx$$e/\Delta$$V_{PG2}$=1 aF. Also the charging energies $E_{C1}$=$\alpha_{1}$$\cdot$$\Delta$$V_{PG1}$=2.57 meV and $E_{C2}$=$\alpha_{2}$$\cdot$$\Delta$$V_{PG2}$=4.77 meV are obtained using the voltage-energy conversion factor $\alpha_{1(2)}$=e$V_{b}$/$\delta$$V_{PG1(2)}$ extracted from the bias triangle as shown in Fig. \ref{Fig3}(e). The difference in charging energies reflects the fact that the sizes of the dots are not equal and can be justified if the tunnel barriers defined by local disorder potential modify the size of GQDs \cite{Liu2010,Volk2011}. Within this picture, electrons from the source reservoir enter through a large localized state in QD$_{1}$ ($E_{C1}$ is small) to a small localized state in QD$_{2}$ ($E_{C2}$ is large), and then exit through the drain reservoir. Finally, the interdot coupling energy can be determined from the splitting of the triangles as shown in Fig. \ref{Fig3}(e): $E_{Cm}$=$\alpha_{1}$$\cdot$$\Delta$$V^{m}_{PG1}$=0.29 meV.

\maketitle
\section{III. Charge stability diagram in perpendicular magnetic field}

\begin{figure*}[!t]	
\includegraphics[scale=1.2]{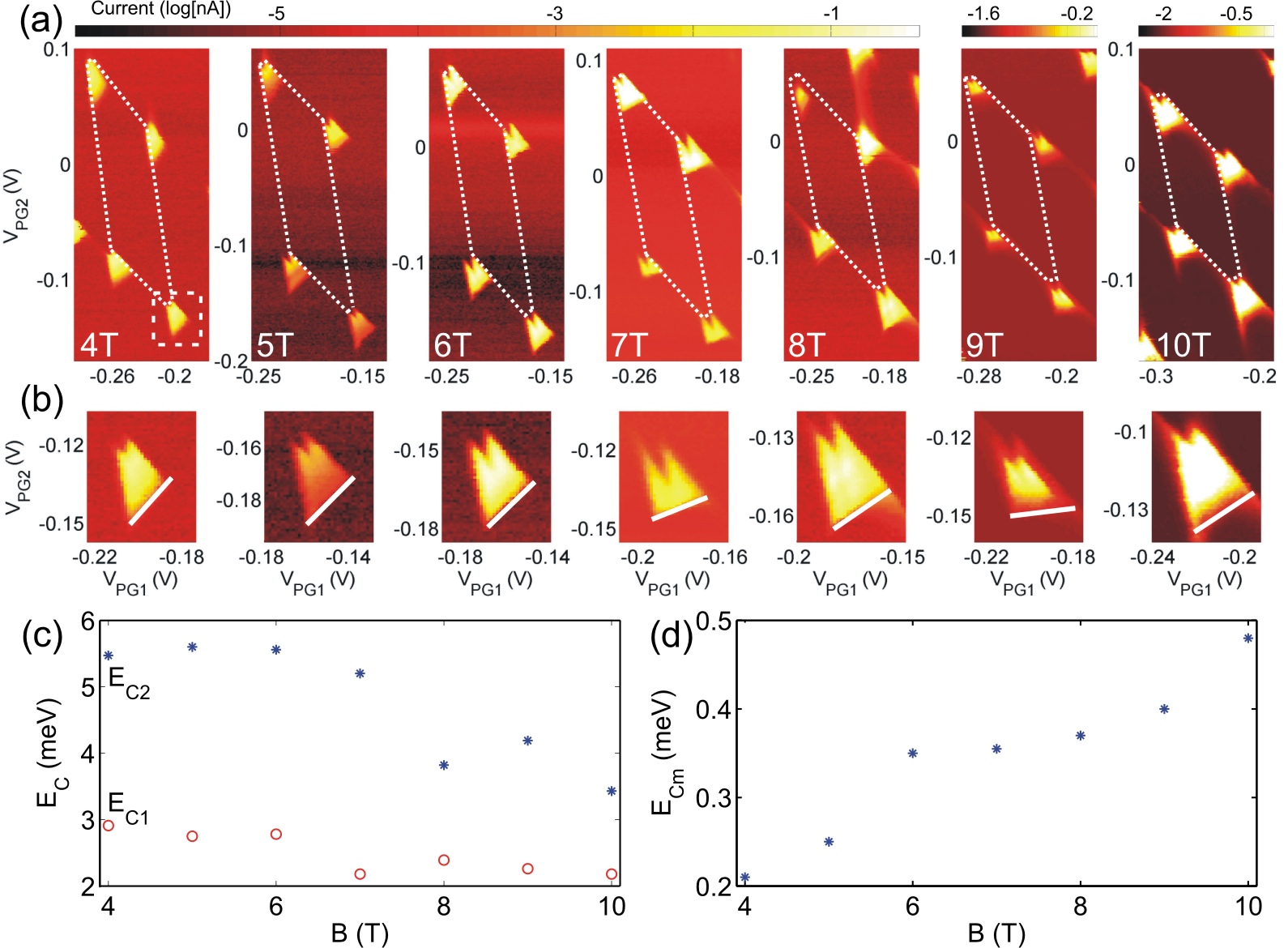}
\caption{(a) The evolution of the charge stability diagram (taken at \textit{T}=100 mK, $V_{BG}$=8.61 V and $V_{SD}$=-1 mV) under the influence of perpendicular magnetic field from 4 to 10 Tesla. (b) Close-up of the triangle in the lowest position of each panel, as highlighted by the dashed square in the leftest panel of (a). (c) The charging energies of the QDs as a function of \textit{B}-field. (d) The interdot coupling energy as a function of \textit{B}-field.}
\label{Fig4}
\end{figure*}

The charge distribution in the QDs can be investigated by looking at how the charge stability diagram evolves under the influence of magnetic field. Fig. \ref{Fig4}(a) shows the evolution of a region of the stability diagram, measured at \textit{T}=100 mK, $V_{BG}$=8.61 V and $V_{b}$=-1 mV, for perpendicular magnetic fields ranging from 4 to 10 T. We have studied the stability diagram in a wide energy range and here only focus on four typical triple points for simplicity. The first observation is the field-dependent change in the dimensions of the honeycomb, which is highlighted by the dotted hexagonal outlines in Fig. \ref{Fig4}(a) and is most pronounced from $B$=7 T to $B$=10 T, indicating the variation in the capacitances $C_{g1}$ and $C_{g2}$. In addition, the close-ups of the triangle in the bottom of each honeycomb, as shown Fig. \ref{Fig4}(b), display a change in $\delta$$V_{PG1(2)}$ and $\Delta$$V^{m}_{PG1(2)}$, implying that the conversion factors $\alpha_{1(2)}$ and interdot coupling energy $E_{Cm}$ also change with \textit{B}-field. It is worth noting that the size of triangle varies in the same honeycomb (i.e., \textit{B}=7 T and \textit{B}=8 T), indicating the precise size of the localized state can change over a small range of gate voltage. The extracted charging energy of each dot ($E_{C1}$ and $E_{C2}$) and the interdot coupling energy ($E_{Cm}$) are shown in Fig. \ref{Fig4}(c) and (d) respectively. The QDs charging energies remain roughly unvaried ($E_{C1}$$\approx$3 meV, $E_{C2}$$\approx$5.5 meV) from \textit{B}=4 T to \textit{B}=6 T then show a decreasing tendency from \textit{B}=7 T to higher fields. From \textit{B}=4 to \textit{B}=10 T, the percentage change in $E_{C2}$ (37.3\%) is larger than that in $E_{C1}$ (25\%). By contrast, the interdot coupling energy shows a monotonic increase with field from 4 T to 10 T. Our results suggest that both dots increase their size at high \textit{B}-fields, which reflects on the decreasing charging energies.

\begin{figure}[!t]	
\includegraphics[scale=0.43]{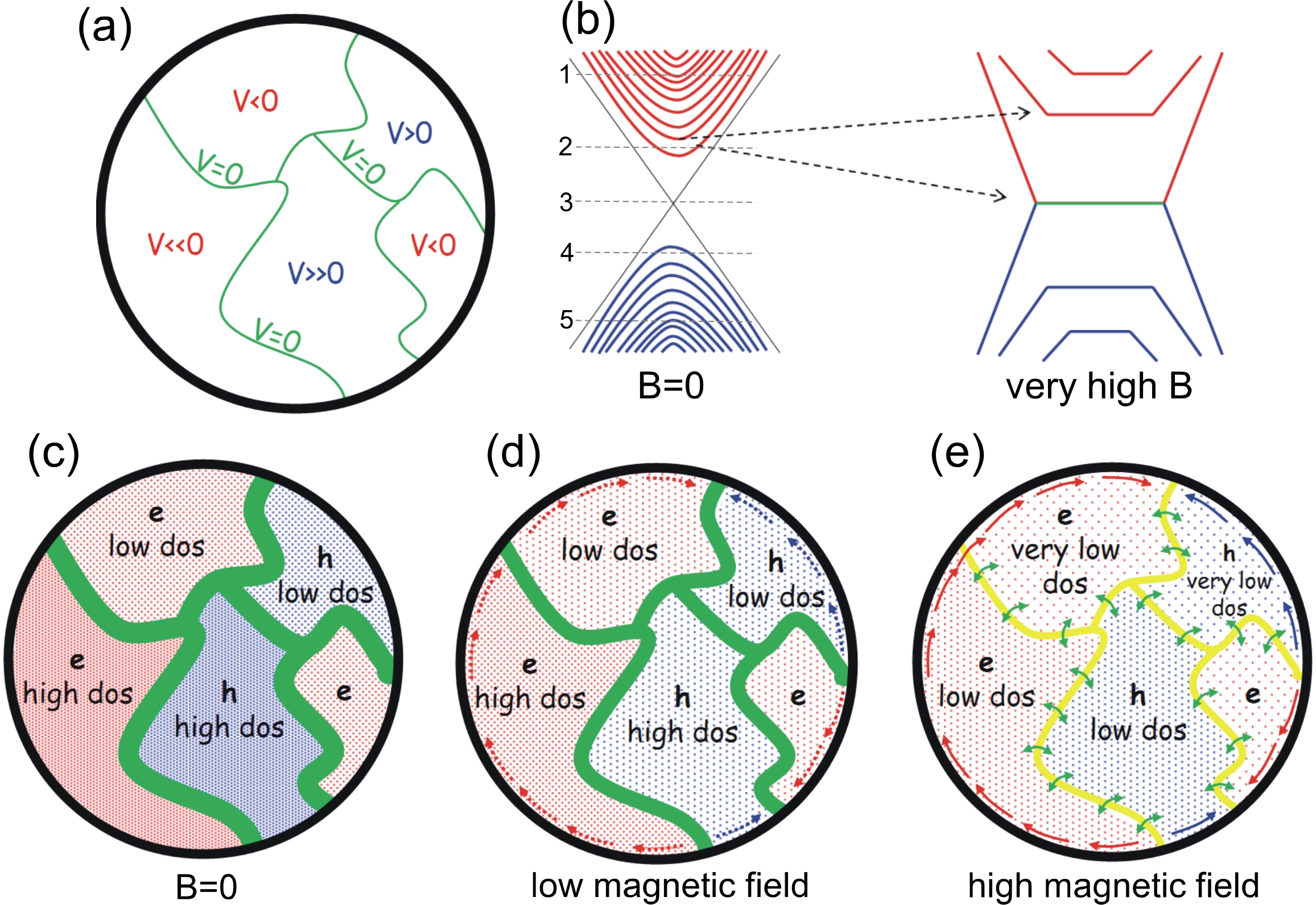}
\caption{(a) Example of potential distribution in a large disordered quantum dot. (b) Left panel: Schematic band structure of a GQD in zero magnetic field. Right panel: in high magnetic field. The red lines denote the electron-like levels and blue lines denote the hole-like levels. The black solid lines indicate the Dirac cone-like dispersion. (c, d, e) Expected DOS distribution in the dot at zero field, low magnetic field and high magnetic field respectively.}    
\label{Fig1}
\end{figure}

\maketitle
\section{IV. MODEL AND SIMULATION}

It is well-known that the presence of charged impurities in the SiO$_{2}$ substrate \cite{Zhang2009,Martin2008} or surface ripples \cite{Deshpande2009} can induce electron-hole puddles with a size of tens of nanometres in exfoliated graphene flakes. This aspect considerably affects the electronic and transport properties of graphene around the Dirac point and, as we will show, it plays a key role in our case. To take this into account, we consider a varying background potential \textit{V} in a model QD, as shown in Fig. \ref{Fig1}(a), where \textit{V} fluctuates from positive (red) to negative (blue) passing from \textit{V}=0 (green). If \textit{V} varies slowly, in each region of a large dot the energy bands will approximately correspond to the shifted energy bands of 2D graphene, as represented in Fig. \ref{Fig1}(b) for zero (left panel) and high (right panel) magnetic field. At \textit{B}=0 T, a gap is introduced to include the quantum confinement effects due to the dot. This gap progressively reduces at high \textit{B}-field along with the formation of Landau levels. Depending on the back-gate and background potentials, the Fermi energy $E_{\rm F}$ (here set to 0) and Dirac point can have locally different relative positions, as indicated by dashed lines in the left panel of Fig. \ref{Fig1}(b). The sign and strength of \textit{V} determine the nature (electron or hole) of the puddles and their density of state (DOS). We first consider the case \textit{B}=0. For the \textit{V}$<<$0 (\textit{V}$>>$0) regions, the Fermi energy corresponds to level 1 (5) in Fig. \ref{Fig1}(b). As the level is far above (below) the charge neutrality point, it gives rise to the electron (hole) puddles with high DOS as shown in Fig. \ref{Fig1}(c). In the region where \textit{V}$<$0 (\textit{V}$>$0), the Fermi energy corresponds to level 2 (4) and results in electron (hole) puddles with low DOS. In the region around \textit{V}=0 corresponding to the energy gap, the DOS is very low or 0. These regions (the green region in Fig. \ref{Fig1}(c)) separate the puddles and can make the transport diffusive \cite{Li2011}.

In the presence of high magnetic fields, due to the formation of LLs, part of the levels around the gap tends to the 0-th Landau level, thus reducing the gap. The other part rises, thus approaching the higher LLs as shown in the right panel of Fig. \ref{Fig1}(b), where also the dispersive magnetic edge states are represented. In the low field regime, the LLs are far from being fully established and the DOS in the dot is low. However, the edge channels start developing with opposite chirality for electron and hole puddles as indicated by arrows in Fig. \ref{Fig1}(d). At high magnetic field, the LL$_{0}$ is well developed with the consequent closing of the band gap. Therefore, in the \textit{V}=0 region the DOS is expected to increase and result in development of non-chiral channels connecting the puddles, as shown in the yellow region in Fig. \ref{Fig1}(e). At the same time, the other LLs start developing together with the chiral magnetic edge channels. In this regime, the DOS decreases in the bulk of the puddles while it increases at their edges. Electron transport through the dot is not confined in a particular puddle but can be delocalized in the dot through flowing in both the chiral edge channels (red or blue arrows) and non-chiral channels (yellow region).

\begin{figure}[!t]	
\includegraphics[scale=0.8]{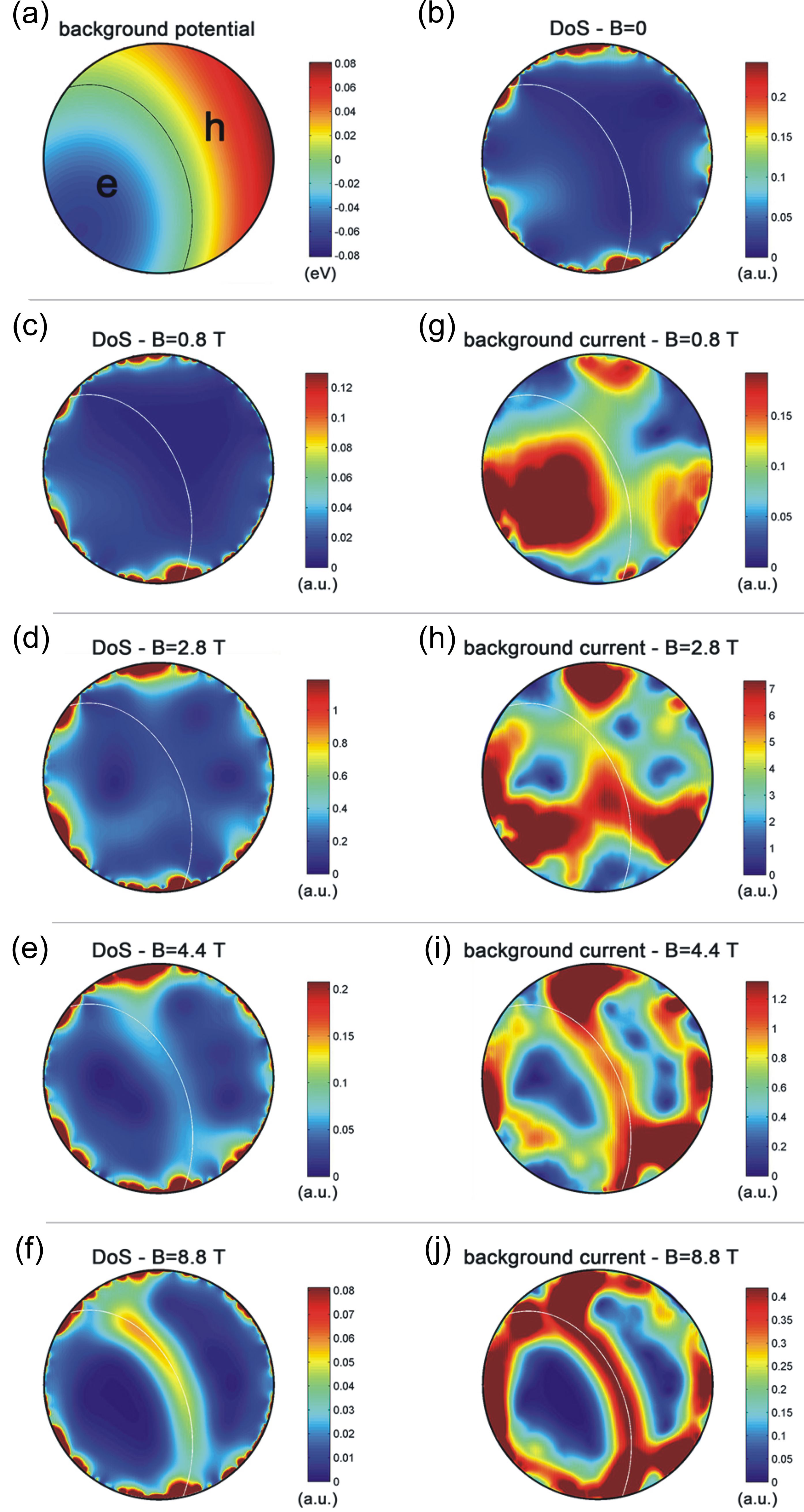}
\caption{(a) Potential distribution in a quantum dot with \textit{R}=47.5 nm. The black line in the potential profile indicates the region where the potential is \textit{V}=0. (b)-(f) Calculated local DOS in the dot at different magnetic field. (g)-(j) Calculated current distribution in the dot at different magnetic field.}    
\label{Fig2}
\end{figure}

In order to validate this picture, we performed numerical simulations of a QD with radius \textit{R}=47.5 nm and a background potential consisting of two regions with \textit{V}$>$0 (81 meV) and \textit{V}$<$0 (-66 meV), which determine the presence of a hole puddle and an electron puddle, as shown in Fig. \ref{Fig2}(a). The dot is described by a first-neighbour tight-binding Hamiltonian with a single $p_z$ orbital \textit{per} atom and coupling parameter -2.7 eV. For the more details on the Green's function formalism adopted for the simulations, refer to Ref. \cite{Cresti2006}. The calculated DOS of the dot is shown in Fig. \ref{Fig2}(b)-(f) in arbitrary units. As expected, at low \textit{B} (0 and 0.8 T) the DOS is low where \textit{V}=0 and higher for larger $|V|$. As \textit{B} increases (2.8 T), the DOS decreases a little in the centre of the electron/hole puddles, and it increases along the edge due to the progressive developing of magnetic edge states (Fig. \ref{Fig2}(d)). Note the presence of very high DOS region at the border of the dot. They correspond to zigzag edge sections, where very localized states appear \cite{Nakada1996}. At higher \textit{B}-field, we observe the presence of high DOS in the \textit{V}=0 region, which corresponds to the LL$_{0}$, and the rise of edge states around the dot. The higher the field is, the larger the DOS is in the \textit{V}=0 and edge regions, as can be seen for \textit{B}$\geq$4.4 T in Fig. \ref{Fig2}(e) and (f).

The simulated background spectral current distribution (which corresponds to the spatial distribution of the conductive channels) \cite{Cresti2006} in the dot is shown in Fig. \ref{Fig2}(g)-(j). At low magnetic field (\textit{B}$\leq$0.8 T) the current is mainly concentrated in the high $|V|$ regions and the \textit{V}=0 region seems to act as a barrier between the two puddles. At slightly higher field (\textit{B}=2.8 T) the current starts tending to the \textit{V}=0 region due to the progressive closing of the energy gap. At high fields (\textit{B}$\geq$4.4 T), we observe the current flowing along the chiral magnetic edge states of the dot and along the non-chiral \textit{V}=0 region, where the gap is now closed. 
In this regime, the current is delocalized in the dot and a charge rearrangemet can be seen compared to the case at low \textit{B}-fields. Note that the more fractured the disorder potential is (meaning more existing electron/hole puddles) the more pronounced the charge delocalization effect will be at high fields. 

\maketitle
\section{V. Discussion}
As the back gate voltage ($V_{BG}$=8.61 V) where all the measurements were carried out is near the charge neutrality point, it is expected the background potential fluctuations will play a role and give rise to electron-hole puddles formed in the QDs \cite{Martin2008,Connolly2010}. In this situation, our model can be readily adapted to explain the data. Due to the closing of the energy gap at high enough magnetic fields, at the \textit{V}=0 region the DOS is high and develops non-chiral channels which connect the puddles, as sketched in Fig. \ref{Fig1}(e). Hence, the current can flow through the puddles \textit{via} crossing the high DOS non-chiral channels at the interface, thus making electrons no longer confined in a particular puddle but delocalized in a larger puddle, resulting in a smaller charging energy. Here we point out that the field has to be high enough for the LLs and edge states to be fully developed to close the energy gap. The threshold \textit{B}-field for this to happen in a GSQD with a relatively smaller size (50 nm diameter) is around 10 T \cite{Guttinger2009}. As a result, the change in charging energies in our case is most pronounced from \textit{B}=6 T to \textit{B}=10 T for QD$_{1}$ and from \textit{B}=7 T to \textit{B}=10 T for QD$_{2}$, as shown in Fig. \ref{Fig4}(c). The threshold \textit{B}-field for QD$_{2}$ is higher owing to its larger charging energy (smaller puddle), in which the magnetic length $\ell_{B}$=$\sqrt{\hbar/eB}$ has to be comparable or even smaller than the puddle size. The magnetic length for \textit{B}=7 T is around 9 nm, implying the size of puddle in the dot is around (or more than) twice the critical magnetic length, in good agreement with the puddle size (20 nm) measured in graphene \cite{Zhang2009}. In addition, we observed the change in $E_{C2}$ from \textit{B}=4 T to \textit{B}= 10 T is larger than that in $E_{C1}$. This is expected, since $E_{C1}$ is smaller than $E_{C2}$, indicating electrons tunnel through a larger localized state in QD$_{1}$ and a smaller localized state in QD$_{2}$. In other words, the disorder potential in QD$_{2}$ is more fractured than that in QD$_{1}$. Therefore, at high \textit{B}-field the charge delocalization effect is more pronounced in QD$_{2}$ than that in QD$_{1}$, giving rise to the larger \textit{B}-dependent charging energy in QD$_{2}$. Here we note that the GQD has to be large for the substrate disorder to play an important role, which may be the reason that the decreasing charging energy with \textit{B}-field is not observed in other relatively smaller GQD with large edge-to-bulk ratio \cite{Gutinger2008,Guttinger2009} or GQD on hBN \cite{Epping2013} where substrate disorder is less important. The increasing interdot coupling energy may be also understood as the charges rearranging from the center of the puddles to the edge of the dot (see Fig. \ref{Fig2} (g)-(j)). This scenario depends on the progressively formed edge state with increasing \textit{B}-field, and can be observed in the whole range of \textit{B}-field (Fig. \ref{Fig4}(d)).

A more recent study of GNR on hBN substrate has indicated that the localized states may also extend into the leads of the device, giving rise to smaller charging energies than expected from the geometry of GNR alone \cite{Bischoff2014a}. However, two conditions are crucial for this effect to be seen. One is the substrate disorder has to be much weaker than the edge disorder, and the other is the edge-to-bulk ratio of device has to be large enough for the edge to play an important role. Transport in GQDs on hBN is dominated by the edge roughness for QDs with diameter less than 100 nm \cite{Engels2013}. Each condition is met by their relatively small GNR (30 nm $\times$ 30 nm) on hBN, thus the edge disorder is strong enough to localize electron wavefunction along the edge to the leads. On the contrary, our large dot (and the tunnel barrier GNR) with smaller edge-to-bulk ratio should diminish the influence of the rough edges on overall transport, meaning localization along the edge still happens but transport is dominated by bulk contributions. Therefore, we argue that the charge redistribution (based on substrate disorder) in our QDs is the main factor that leads to a variation of dots area and contribute to the decreasing charging energies in magnetic fields. 

\begin{figure}[!t]	
\includegraphics{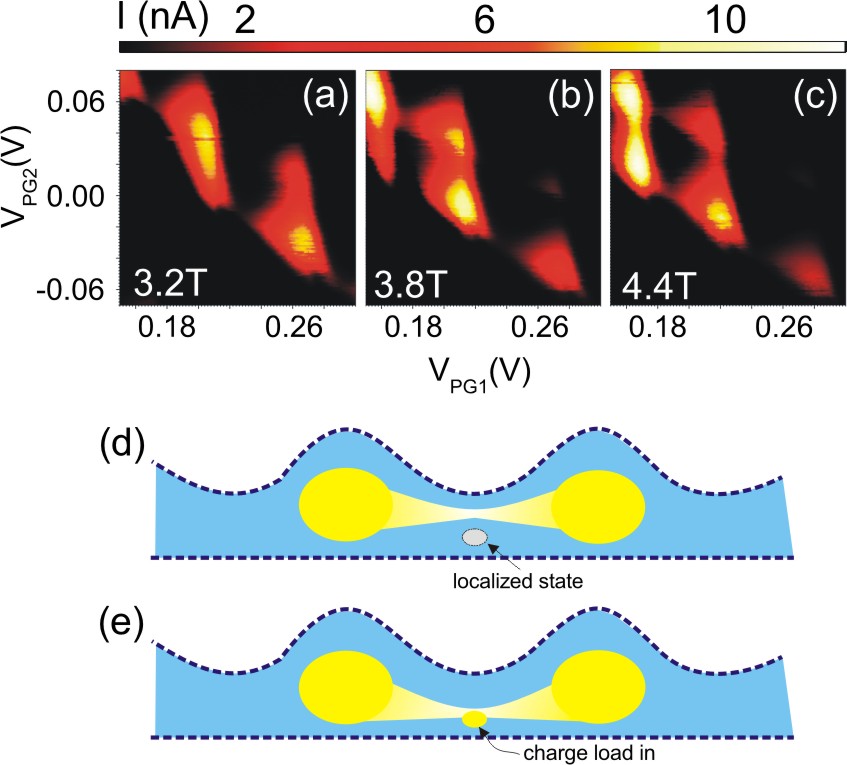}
\caption{The charge stability diagram measured at $V_{BG}$=9 V and $V_{b}$=1 mV in (a) \textit{B}=3.2 T, (b) \textit{B}=3.8 T and (c) \textit{B}=4.4 T showing a formation of an additional dot under the influence of magnetic field and strong couple to the original dots. (d) Graphic illustration of an effect of a localized state formed in magnetic fields. (e) Same as (d) but with a charge added into the localized state. It capacitively couples to the original dots and forces the DQD to reconstruct its wavefunction.}
\label{Fig5}
\end{figure}

The effect of disorder can be also seen in Fig. \ref{Fig5}, where a stability diagram measured in different cool-down of device at \textit{B}-fields (a) 3.2 T, (b) 3.8 T and (c) 4.4 T is presented. The triangle shape first distorts at \textit{B}=3.2 T and then splits into two separated ones (Fig. \ref{Fig5} (b)), then moves further apart and form an additional row of triangles (Fig. \ref{Fig5} (c)). We attribute this newly appeared triangles to the formation of a localized state in a magnetic field, which is capacitively coupled to the original dots \cite{Guttinger2011,Wei2013}. A schematic illustration is shown in Fig. \ref{Fig5} (d) and (e) to address such a scenario. When a localized state is formed in magnetic field, while the gate voltage is sweeped it can add or subtract charges discretely to the parasitic dot, thus altering the entire environment abruptly and unexpectedly. The fact that the splitting occurs on both gate spaces suggests the localized state can affect two dots in a similar way, implying its location is in the central GNR (Fig. \ref{Fig5} (d)). The new dot acts as a gate which will shift the triple-points in charge stability diagram, consequently, leading to an additional row of triangles adding adjacent to the original ones.

\maketitle
\section{VI. Summary}
In summary, we have fabricated and studied the magneto-transport properties of a large GDQD device. In different cool downs, we observed a honeycomb pattern which is typical of charge stability diagrams for a DQD system. We studied the evolution of the charge stability diagram under the influence of \textit{B}-field up to 10 T. The charging energy and the interdot coupling energy show different dependence with \textit{B}-field, suggesting the size of both dots become larger in high field. Our interpretation is supported by numerical simulations in which we show the confined charges in the puddles of GQDs can be redistributed from the bulk to the edge through the formation of LLs and edge states. At high enough \textit{B}-field, due to the closing of energy gap, electrons are delocalized \textit{via} crossing the non-chiral channels connecting different puddles, resulting in a smaller charging energy. 

This work was financially supported by the EPSRC. Additional data related to this publication is available at the xxxxxxx data repository.

\end{document}